\documentclass[sn-mathphys-num]{sn-jnl}

\usepackage{graphicx}%
\usepackage{multirow}%
\usepackage{amsmath,amssymb,amsfonts}%
\usepackage{mathrsfs}%
\usepackage[title]{appendix}%
\usepackage{xcolor}%
\usepackage{textcomp}%
\usepackage{manyfoot}%
\usepackage{booktabs}%
\usepackage{algorithm}%
\usepackage{algorithmicx}%
\usepackage{algpseudocode}%
\usepackage{listings}%

\graphicspath{{figures/}{./}}

\begin{document}

\title[Collective Motion of Pedestrians in a Three-Way Crossing]
{Interacting Streams of Cognitive Active Agents in a Three-Way Intersection}


\author[1]{\fnm{Priyanka} \sur{Iyer}}\email{p.iyer@fz-juelich.de}

\author[1]{\fnm{Rajendra Singh} \sur{Negi}}\email{r.negi@fz-juelich.de}

\author[2]{\fnm{Andreas} \sur{Schadschneider}}\email{as@thp.uni-koeln.de}

\author*[1,2]{\fnm{Gerhard} \sur{Gompper}}\email{g.gompper@fz-juelich.de}

\affil*[1]{\orgdiv{Theoretical Physics of Living Matter, Institute of Biological Information Processing and Institute for Advanced Simulation}, \orgname{Forschungszentrum J{\"u}lich}, \orgaddress{\city{J{\"u}lich}, \postcode{52425}, \country{Germany}}}

\affil[2]{\orgdiv{Institute for Theoretical Physics}, \orgname{Universit\"at zu K\"oln}, \orgaddress{\city{K\"oln}, \postcode{10587}, \country{Germany}}}


\abstract{The emergent collective motion of active agents - in particular pedestrians - at a three-way 
intersection is studied by Langevin simulations of cognitive intelligent active Brownian particles (iABPs) with directed visual perception and self-steering avoidance. Depending on the maneuverability $\Omega$, the goal fixation $K$, and the vision angle $\psi$, different types of pedestrian motion emerge. At intermediate relative maneuverability $\Delta=\Omega/K$ and large $\psi$, 
pedestrians have noisy trajectories due to multiple scattering events as they 
encounter other pedestrians in their field of view. For $\psi=\pi$ and 
large maneuverability $\Delta$, an effectively jammed 
state is found, which belongs to the percolation universality class. For small 
$\psi$, agents exhibit localised clustering 
and flocking, while for intermediate $\psi$ self-organized rotational flows can emerge. Analysing the mean squared displacement and velocity 
auto-correlation of the agents revealed that the motion is well described by fractional Brownian Motion with positively correlated noise.  Finally, despite the rich variety of collective behaviour, the fundamental flow diagram for the three-way-crossing setup shows a universal curve for the different 
vision angles.  Our research provides valuable insights into the importance of 
vision angle and self-steering avoidance on pedestrian dynamics in semi-dense 
crowds.}




\maketitle

The understanding of collective pedestrian movement is imperative for the design 
of strategies facilitating smooth pedestrian flow in crowded areas, the mitigation 
of crowd-related disasters in confined spaces, and the development of evacuation 
procedures \cite{FelicianiBook}. Here, an important aspect is the goal-oriented
motion of all participants. For pedestrian navigation in crowds, typical
scenarios are the formation of traffic jams in front of narrow passages and bottlenecks, 
the interaction of groups in counter flow leading to lane formation, 
and the self-organization of flows at intersections (see the reviews \cite{SchadschneiderCSTZ18,corbetta2023physics} and references therein).
Situations like the Shibuya Crossing in Tokyo or mall intersections pose important questions regarding self-organization and the design optimization of facilities. Experiments and simulations of bi-directional flows demonstrate lane formation \cite{ZhangKSS12,FelicianiN16b,dong2017analysis}, while cross flows at an angle result in stripe-like patterns \cite{mullick2022analysis}. 
Four-directional cross-flow experiments 
\cite{BodeCH19,cao2017fundamental,lian2015experimental,SunHGQC18} and multi-directional 
crossing scenarios explored through circle antipode experiments \cite{xiao2019investigation,hu2019experimental}, with participants positioned on a circle and crossing diagonally, have been used to study navigation strategies, conflict avoidance etc.

The importance of understanding the flow of pedestrian crowds has led to the 
development of various modelling approaches in recent decades, e.g. the force-based models, cellular automaton-based approaches, several physics-inspired models, game theory, optimal control and fluid dynamics (for a more detailed exposition of the different approaches, see \cite{MartinezGil2017,MauryBook,SchadschneiderCSTZ18,corbetta2023physics} and references therein). The importance of a close interplay between empirical and theoretical investigations has inspired the use of specifically designed laboratory experiments \cite{BoltesZS13,SchadschneiderCSTZ18,corbetta2023physics}, which generate quantitative results that are important benchmarks for modeling. 

Collective pedestrian motion can be generically understood as the behavior of 
self-propelled interacting entities, which places it into the realm of the large
field of ``active matter'' \cite{gompper20202020}, which encompasses systems from 
suspensions cells and self-propelling colloids to schools of fish and flocks of 
birds. In this context, the active Brownian Particle (ABP) model has been used 
extensively to understand many intriguing aspects of non-equilibrium physics such 
as mobility-induced phase separation \cite{fily2014dynamics,cates2015motility} and
wall accumulation \cite{iyer2023dynamics,caprini2022dynamics}. 
Moreover, when equipped with directional environment sensing and self-steering, 
ensembles of 'intelligent' ABP systems (iABPs) can show a rich variety of collective 
phenomena such as milling, 
single-file motion, flocking, worm-like swarms, and polar or nematic ordering 
\cite{Couzin_2005_nature, barberis_2016_PRL, negi2022emergent, negi2024collective}. 
In pedestrian models, vision-based sensing and cognitive steering are also key 
ingredients that determine the emergent collective behaviour 
\cite{ondvrej2010synthetic,wirth2023neighborhood,zhang2022hdrlm3d,dachner2022visual}, 
which suggests the applicability of iABP models for the description of pedestrian 
crowds \cite{negi2024controlling}.

\begin{figure*}
    \centering
    \includegraphics[scale=0.5]{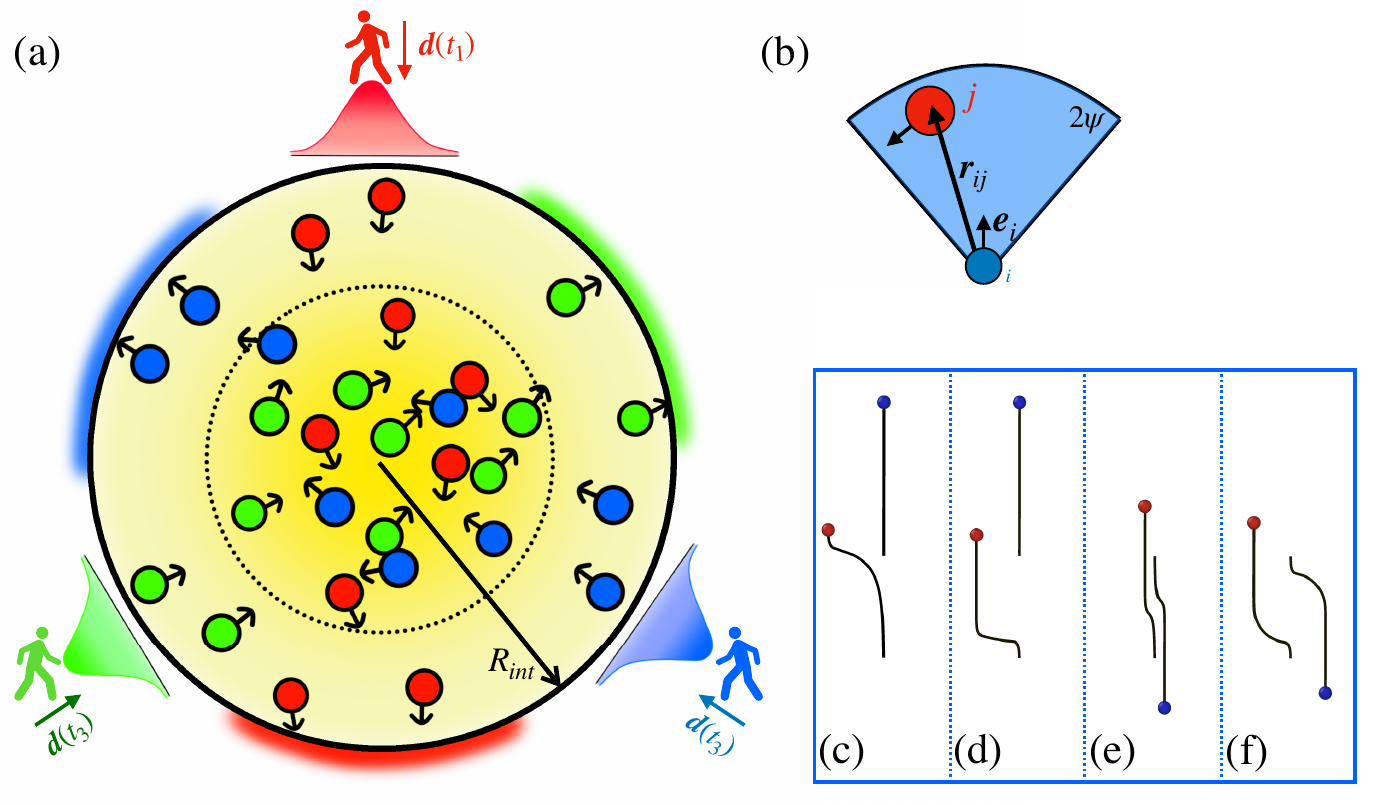}
    \caption{(a) Simulation setup of a three-way pedestrian crossing. The colours represent different pedestrian types $t_i$, i.e. pedestrians with different goals [alignment along $\boldsymbol{d}(t_i)$]. The human markers show the position of influx, each separated by an angle of $\pi/3$ from the other and placed on an interaction circle of radius $R_\text{int}=120R_0$. The shaded regions depict the regions of 'successful exits'[see Supplementary Information (SI) \cite{supp}], and the distribution at the inflow indicates the 'spread' of each stream leading to an effective interaction zone (dashed circle) $R_\text{int}/2$ [see Methods]. (b) Schematic diagram of the vision-based interaction of agent $i$ with agent $j$. The vision angle is highlighted in blue, with a vision angle $\psi$ and cutoff $4R_0$.  Sample trajectories showing the effect of alignment and visual avoidance for a vision angle of $\psi=\pi/2$. Agents with the same goal direction for (c) $\Delta=1$ and (d) $\Delta=2$. The blue agent does not 'see' the red one and therefore does not react. Agents with opposite goal directions for (e) $\Delta=1$ and (f) $\Delta=2$. In this case, both agents see each other and move away. In all cases at $t=0$, the distance between the agents is $r=3R_0$.  }
    \label{fig:schematic}
\end{figure*}

{\em Model and Cross-Stream Setup --} We investigate here a three-stream 
intersection scenario [see Fig.~\ref{fig:schematic}(a)], 
which emulates a basic realization of multi-directional flows in a circle. In contrast to a straightforward two-way flow configuration, pedestrian movement at intersections with multiple streams does not readily organize itself through lane or stripe formation, making it an important case to study. A similar setup with two intersecting streams has been studied experimentally \cite{BodeCH19}. Since we are interested 
in the general physical mechanisms of interacting streams, sophisticated models,
which usually have several adjustable parameters, are not appropriate. Instead, 
pedestrians are modelled as intelligent active Brownian particles (iABPs) in two
spatial dimensions [see Fig.~\ref{fig:schematic}(b), which experience a propulsion 
force $f_p$ acting along their orientation vector $\mathbf{e}_i$, and a friction 
force $-\gamma \mathbf{v}_i$ with velocity $\mathbf{v}_i$, which implies a
constant speed $v_0 = f_p/\gamma$. Each pedestrian is associated with a type $t_i$, which encodes their goal direction $\hat{\boldsymbol{d}}(t_i)$. We employ a self-steering mechanism in the form of a torque that changes the direction of motion as
\begin{equation}
  \dot{\mathbf{e}}_i =\sqrt{2(d-1)D_r}\boldsymbol{\Lambda}_i 
       + \Omega \mathbf{M}_\text{vis} + K \mathbf{M}_\text{goal}, 
    \label{eq:steering}
\end{equation}
where $D_r$ is the rotational diffusion coefficient, $d$ is the dimensionality, $\boldsymbol{\Lambda}_i$ 
is a Gaussian random process, $\Omega$ and $K$ are the strength of the vision ($\mathbf{M}_\text{vis}$) 
and goal-fixation ($\mathbf{M}_\text{goal}$) steering torques, respectively [see Methods for details]. The vision steering torque $\mathbf{M}_\text{vis}$ aligns the orientation vector $\mathbf{e}$ away from the center of mass of agents in the vision cone, while the goal-fixation torque $\mathbf{M}_\text{goal}$ aligns $\mathbf{e}$ with the goal vector $\hat{\boldsymbol{d}}(t_i)$, see Fig.~\ref{fig:schematic}(a). The vision-based steering torque also contains a weight factor that increases the relative importance of avoiding agents moving head-on toward each other by a factor $1/2$ relative to co-moving agents \cite{karamouzas2014universal}. The activity of the agents is described by the dimensionless P{\'e}clet number 
$\text{Pe} = f_p/(\gamma R_0 D_r)=v_0\tau_r/R_0$, where $\tau_r=1/D_r$ is the 
rotational diffusion time, $v_0=f_p/\gamma$ is the agent velocity, and $R_0$ is 
the effective vision range. The system is studied for varying relative 
maneuverability $\Delta=\Omega/K$, vision angle $\psi$, and inflow rate $\Gamma$. 
We operate in the limit of over-damped motion, so that inertial effects are 
negligible and the self-steering gives a realistic description of pedestrian 
cognitive motion, see Fig.~\ref{fig:schematic}(c-f). This also avoids the 
conceptual problems of forced-based models which are mostly a consequence of 
strong inertia effects (see e.g.\  Ref.~\cite{CordesCTS23}). The agents are 
considered point particles, i.e.\ the simulations are in the limit of 
semi-dense crowds, where the volume-exclusion radius $\sigma$ of an individual pedestrian is much smaller than the vision range $R_o$, i.e.\ $\sigma \ll R_0$.

\begin{figure*}
    \centering
    \includegraphics[scale=0.5]{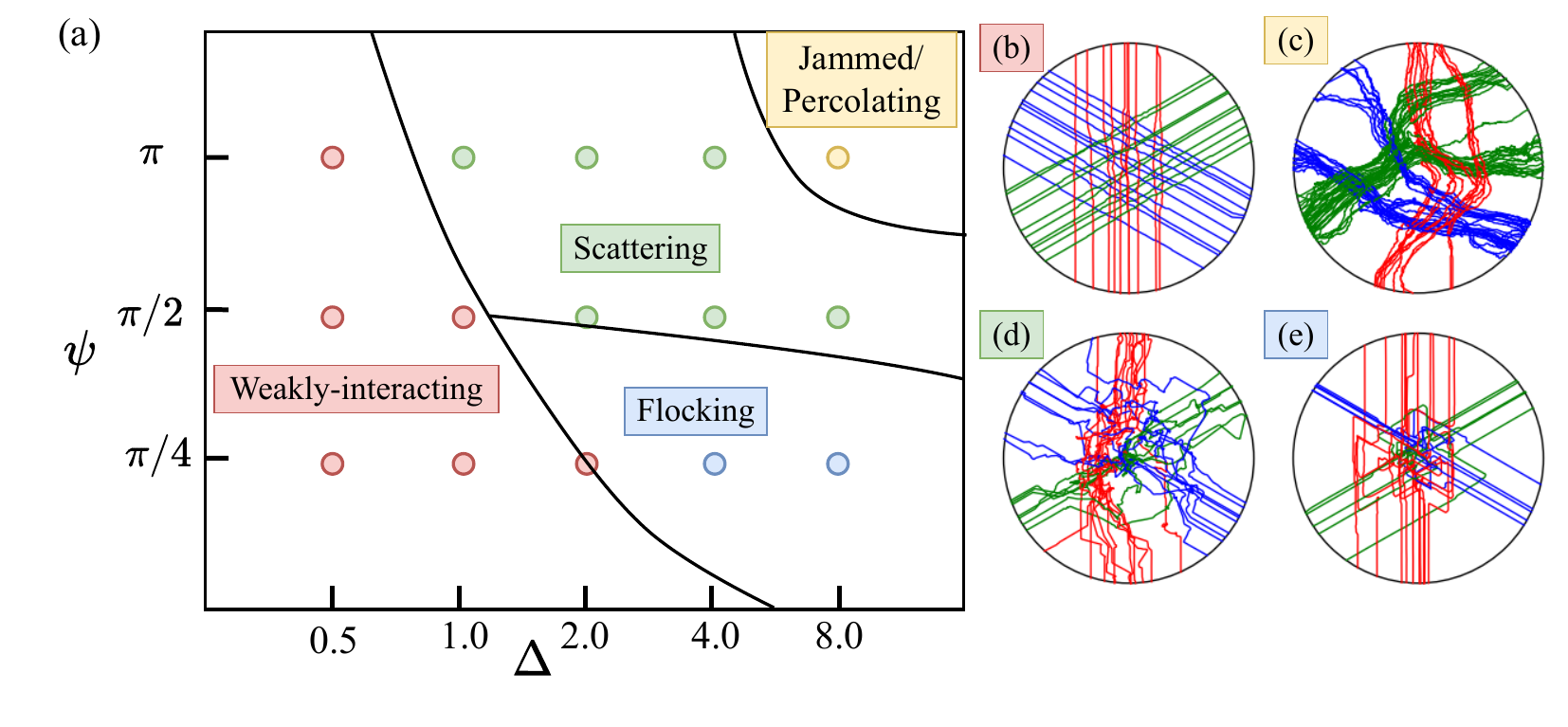}
    \caption{(a) State diagram of pedestrian movement states as a function of the relative  maneuverability $\Delta=\Omega/K$ and vision angle $\psi$. 
    (b) For small $\Delta$, the agents do not avoid each other significantly, and 
    pass through the interaction zone nearly unhindered.  
    (c) For higher $\Delta$, and the largest vision angle $\psi=\pi$ and relative 
    maneuverability $\Delta=8$, a jammed, percolating phase develops. 
    (d) For intermediate $\Delta$ and vision angles $\psi\geq\pi/2$, a scattering
    regime emerges as the agents attempt to avoid each other while crossing. 
    (e) For intermediate $\Delta$ and smaller vision angles $\psi<\pi/2$, 
    a 'flocking' regime is found, where agents navigate by aligning with oncoming 
    individuals, forming a local co-moving pedestrian cluster, thus leading to 
    clustering and flocking -- as seen by the emergence of parallel trajectories 
    (with different pedestrian types). Here we fix $\Gamma=1$.}
    \label{fig:state_diagram}
\end{figure*}

\begin{figure}
    \centering
    \includegraphics[scale=0.7]{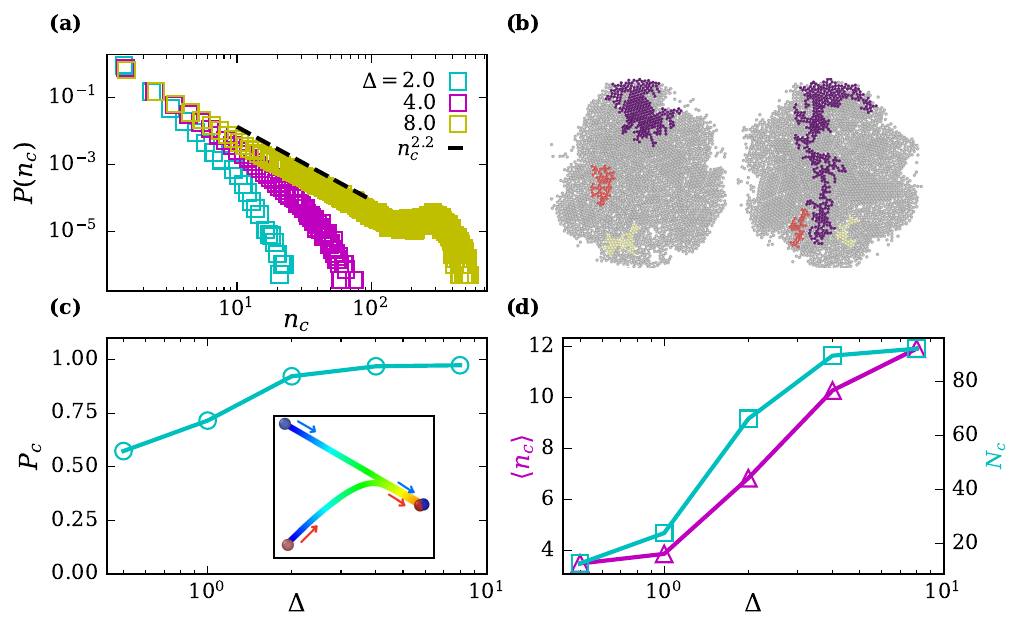}
    \caption{(a) A phase transition into the percolation state occurs as $\Delta$ 
    is increased 
    for vision angle $\psi=\pi$, marked by the development of a power law decay of 
    the cluster size distribution $p(n_c)$ and the large percolated cluster The peak at large cluster sizes is due to 
    finite system size.
    (b) The three largest clusters at $\Delta=8.0$ and $\psi=\pi$ at two different 
    times for pedestrians of one type. 
    (Left) Clusters are dispersed and the largest cluster is at the inflow. 
    (Right) The largest cluster can feed smaller clusters and reach up-to the exit,
    forming a transiently percolated cluster. 
    (c) Average cluster polarization $P_c$ [see Eq. S3 in SI] for $\psi=\pi/4$ shows an increase 
    with $\Delta$, indicating the development of avoidance-induced flocking. 
    The inset shows the trajectory of two agents exhibiting avoidance-based flocking. 
    (d) The transition into the flocking phase is characterized by a strong 
    increase in both the mean cluster size $\langle n_c \rangle$ and the number 
    $N_c$ of clusters. The distance cutoff $R_{cut}$ are chosen to be $R_{cut}\simeq R_v$ and $R_{cut}\simeq R_0$ for vision angles $\psi=\pi$ and $\psi=\pi/4$, respectively [SI]. }
    \label{fig:percolation_flock}
\end{figure}

\begin{figure}
    \centering
    \includegraphics[scale=0.7]{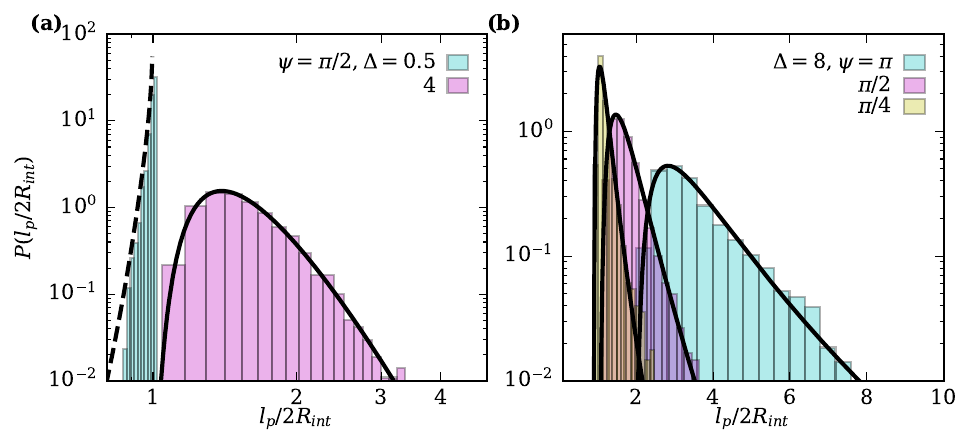}
    \caption{Probability distribution $P$ of the path length $l_p$ for (a) different 
    relative maneuverability ($\psi=\pi/2$, $\text{Pe}=100$) and (b) different vision
    angles ($\Delta=8$, $\text{Pe} = 100$). For small $\Delta$, the paths are nearly straight and the probability distribution is well approximated by 
    Eq.~\eqref{eq:time_dist}[dashed line]. In scenarios characterized by high 
    maneuverability ($\Delta$) and large vision angles ($\psi$), agents traverse 
    longer paths within the interaction sphere to navigate around others. This 
    behavior yields a log-normal distribution for the path length, with the black 
    solid line representing a fitted log-normal model to the data. The path lengths 
    are only determined for trajectories that successfully reach the exit 
    (SI) and are averaged over different agent types. The data is 
    collected for times $4t_0<t<16t_0$ and averaged over all agents, 
    where $t_0=2R_\text{int}/v_0$. }
    \label{fig:angles}
\end{figure}

\begin{figure*}
    \centering
    \includegraphics[scale=0.7]{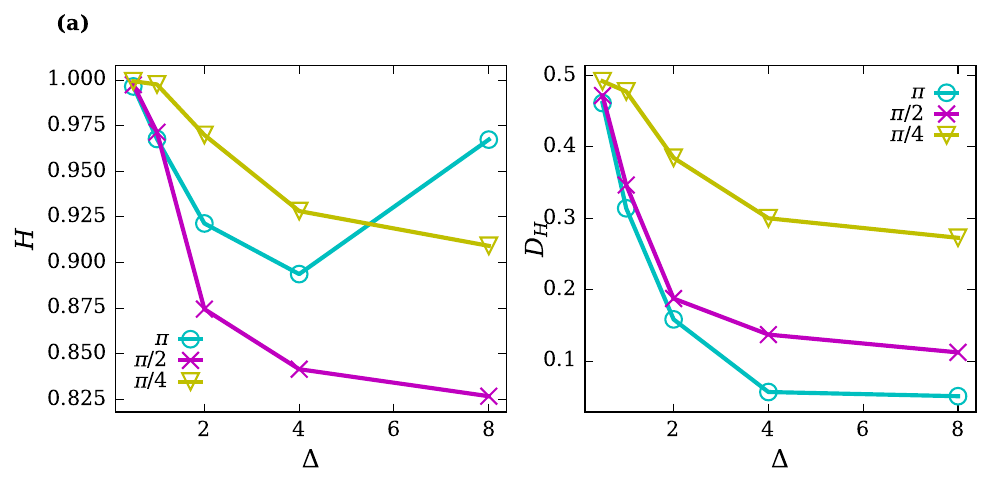}
    \caption{The (a) Hurst exponent $H$ and (b) diffusion coefficient $D_H$ for increasing $\Delta$ and different vision angles $\psi$. The values are estimated by averaging over the two values measured by fitting Eq.~\eqref{eg:corr} and Eq.~\eqref{eq:msd} to C(t) and MSD respectively, and averaging the values over different Pe numbers. Only trajectories that successfully reach the exit are considered and we average over all agents.}
    \label{fig:dynamics}
\end{figure*}

\begin{figure*}
    \centering
    \includegraphics[scale=0.7]{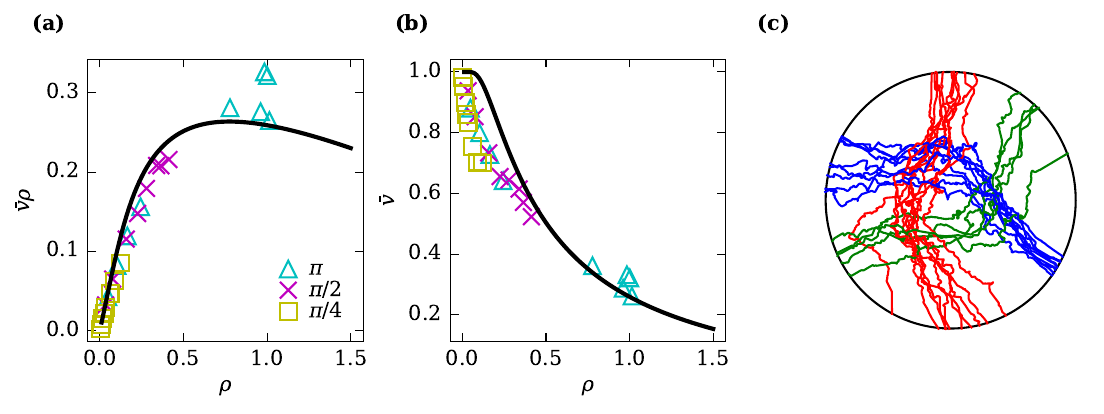}
    \caption{ Fundamental diagrams of the pedestrian flow measured by performing simulations with different pedestrian inflow $\Gamma$ for a fixed $\Delta=8$. (a) Flux $J=\Bar{v}\rho$ and (b) average velocity $\Bar{v}$ as a function of the mean agent density $\rho$. The data collapses onto a single master curve for different vision angles and exhibits the characteristic shape of the fundamental diagram, showing the free flow ($\rho<0.5)$ and jammed regimes ($\rho > 1$). (c) Agent trajectories for $\Gamma=4$ and $\psi=\pi/2$ showing the development of rotational flows and roundabout traffic like motion.  For $\psi=\pi$, a jamming transition occurs for large inflow $\Gamma>0.5$, marked by a sudden rise in density (see $\rho>1.0$) and a strong reduction in velocity $\Bar{v}$. The solid black line in (a,b) is an approximate fit based on the Kladek formula $v(\rho)=v_0[1-\exp({-c[\rho^{-1}-\rho_{jam}^{-1}]})$, where $v_0=1$ and we set $\rho_{jam}=4$ and $c=0.4$ for a good fit.}
    \label{fig:rate}
\end{figure*}

{\em Dynamic State Diagram --} The state diagram of the various collective pedestrian 
movement states as function of relative maneuverability $\Delta$ and vision angle 
$\psi$ is shown in Fig.~\ref{fig:state_diagram}(a). At small $\Delta \lesssim 1$, 
agents essentially ignore each other and head directly toward the goal, see 
Fig.~\ref{fig:state_diagram}(b). This is qualitatively similar to the case of 
panicking pedestrian crowds, where the 'goal' becomes more important than 
avoidance \cite{helbing2001self}. This behaviour corresponds to the case of 'dumb' 
active Brownian particles, and is expected to display activity-induced jamming 
in the presence of excluded volume effects. Thus, efficient navigation requires 
pedestrians to have maneuverability $\Omega$ larger than goal fixation $K$, 
i.e. $\Delta=\Omega/K \gtrsim 1$. The required relative maneuverability $\Delta$ 
increases with decreasing vision angle, as agents see fewer other agents for 
smaller $\psi$. As $\Delta$ increases for larger vision angles, the pedestrian streams start to avoid each other, which results in a complex motion marked by many scattering events, see Fig.~\ref{fig:state_diagram}(d) and Movie M1. While for two intersecting streams, lane (or stripe) formation occurs \cite{helbing2005self}, for three streams the scenario is much more complex and no stable global order exists \cite{BodeCH19}. The agents rapidly change their direction attempting to avoid other agents leading to noisy and convoluted trajectories. For $\psi=\pi$, agents enter a jammed-percolating state, wherein strong clustering is observed and agents cross the interaction regime in groups, see the 'clustered' trajectories in Fig.~\ref{fig:state_diagram}(c).

As the vision angle is further decreased, to $\psi < \pi/2$, the particle motion 
drastically changes, see Fig.~\ref{fig:state_diagram}(e). In this regime, agents 
mainly avoid other agents directly ahead of them, implying that their direction of 
motion changes only for high particle densities, i.e. close to the centre of the 
interaction zone and near the other incoming pedestrian streams. This state is 
characterized by the presence of parallel trajectories in the interaction zone 
[see Fig.~\ref{fig:state_diagram}(e)]. 
Here an agent of one type initially adopts a strategy of polar alignment with the 
oncoming agents of the other types to avoid "collisions". The small vision angle 
is responsible for this flocking-based avoidance mechanism, and has been shown 
recently in Ref.~\cite{negi2024controlling}. No pronounced differences are observed for various 
choices of Pe across all vision angles, which is due to the large goal-fixation ($K/D_r=8$).

{\em Cluster-Size Distributions: Percolation and flocking -- }
As $\Delta$ increases for vision angle $\psi=\pi$, the system undergoes a jamming 
transition due to increased avoidance between agents. Note that the jamming here is
\textbf{\em not} due to volume exclusion, but due to the strong tendency to maintain
a large inter-agent distance in all directions. In the jammed state, the agents 
crowd the interaction regime and form large clusters comprised of agents with
the same goal (or type). Clustering is initiated at the inflow; the clusters then 
extend deep into the interaction region as agents navigate toward their respective 
goals. Remarkably, the jammed state also exhibits percolation, i.e. the clusters
span the length of the interaction zone, see Fig.~\ref{fig:percolation_flock}(a). 
The cluster-size distribution shows a power-law decay, with an exponent 2.2, consistent the percolation universality class \cite{stauffer2018introduction}. 
Therefore, despite of the complex motion and continuously varying environment, the 
movement from the inflow to the exit can be understood qualitatively under the 
realm of percolation theory. The broad peak in the cluster-size distribution at 
$n_c \gtrsim 200$ represents the cluster formed initially at the inflow that 
then feeds smaller clusters into the system which make their way to the exit 
($n_c<100$). Note that the percolation in this case is dynamic, i.e. the system shows transient periods of percolating clusters, interrupted by times when the clusters are dispersed, see Fig.~\ref{fig:percolation_flock}(b).

In the regime of small vision angles, specifically for $\psi=\pi/4$, agents exhibit 
avoidance-induced flocking behavior. Consider two agents moving toward each other 
at a small angle, so that only one of them is visible to the other. The 'aware' 
agent initiates a (slight) turn to avoid a collision. However, the unaffected 
motion of the other agent causes it to repeatedly enter the vision cone of the 
'aware' agent. Consequently, the aware agent must keep turning away until the 
other agent is no longer visible to it. This only happens when they move essentially
parallel to each other, resulting in the formation of a co-moving cluster, as 
illustrated in the inset of Fig.~\ref{fig:percolation_flock}(c). This process repeats when this mini-cluster encounters other agents, who may also align to 
avoid collision, thereby also becoming part of the cluster [see Movie M2]. A particle can only 
leave the cluster if a strong fluctuation disrupts its aligned state. Consequently, 
an avoidance-induced clustering and flocking state emerges as the strength of 
relative maneuverability $\Delta$ increases.

This phenomenon is characterized in Fig.~\ref{fig:percolation_flock}(c,d), where 
a significant increase in the average cluster size $\langle n_c \rangle$, number 
$N_c$ of clusters, and cluster polarization $P_c$ is observed with increasing $\Delta$. For small 
$\Delta \lesssim 1$, polarization remains close to 0.5, indicating a non-flocking 
state. This occurs due to the random overlap of particles from neighboring streams 
that do not avoid each other due to the low self-steering avoidance. However, as 
$\Delta$ increases, the clusters achieve a polarization value near unity, signalling the emergence of a flocking/clustering state. 

{\em Path Length Distributions -- }With increasing relative maneuverability $\Delta$, strong avoidance between agents leads to scattering, and implies larger exit times and broader path-length distributions, as presented in Fig.~\ref{fig:angles} for various $\Delta$ and $\psi$. For $\Delta \lesssim 1$, inter-agent interactions are small, and the path length distribution of nearly straight paths can be estimated to be [SI]
\begin{equation}
     f_L(\Tilde{l_p}) =  \frac{2\Tilde{l_p} \exp[{-(1-(\Tilde{l_p})^2)/2(\Tilde{\sigma})^2}]} {\Tilde{\sigma}\sqrt{2\pi(1-(\Tilde{l_p})^2)}} ,
     \label{eq:time_dist}
\end{equation}
where $\Tilde{l_p}=l_p/2R_\text{int}$ and $\Tilde{\sigma}=\sigma/R_0$ is the normalized 
variance at the input of the stream. For low $\Delta$, the data matches well with 
the estimated distribution, see Fig.~\ref{fig:angles}(a). However, as $\Delta$ 
increases, the distribution shifts to larger $l_p$ and broadens. This occurs as 
agents scatter strongly due to high avoidance $\Delta$, leading to longer paths. 
Agents with lower vision angles reach their destinations in shorter paths, due to 
fewer scattering events as seen in Fig.~\ref{fig:angles}(b), where increasing $\psi$ 
causes a shift of the distribution to larger $l_p$, along with the development of a 
longer tail. The distributions for $\Delta \gtrsim 1$ and $\psi\geq\pi/2$ follow a 
log-normal distribution, which is verified by performing a Kolmogorov–Smirnov test 
with  confidence interval of 95\%. Notably, log-normally distributed path lengths 
have been documented in antipode experiments involving pedestrians initiated on 
a circle \cite{xiao2019investigation}. The experimental arrangement closely mirrors 
the three-stream configuration utilized in our simulations, providing empirical 
support for the shape of the observed distribution.

\begin{figure}[h]
    \centering
    \includegraphics[scale=0.7]{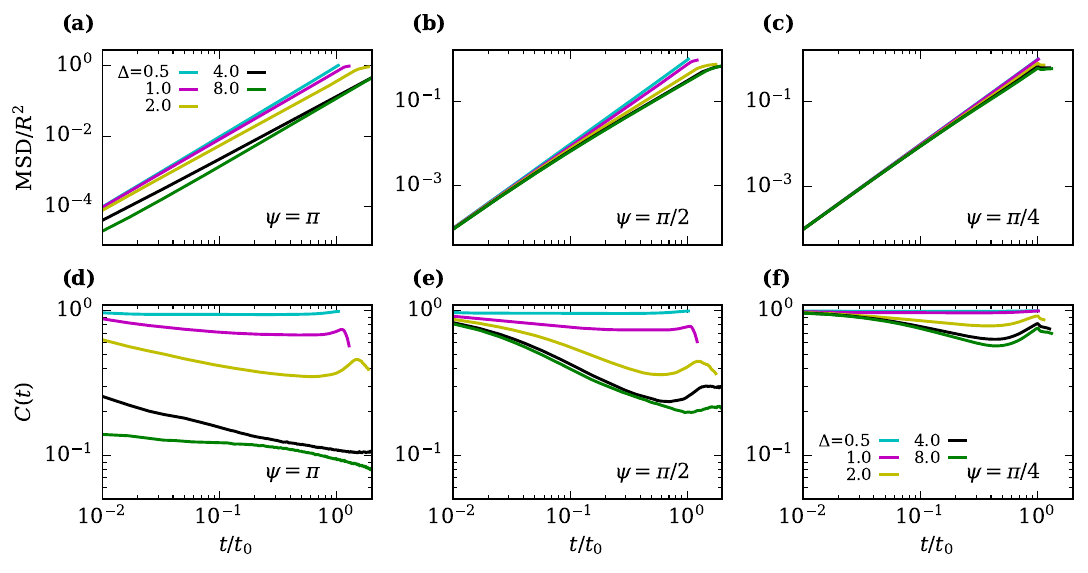}
    \caption{Mean squared displacement (a-c) and the corresponding orientational auto-correlation function (d-f) for various $\Delta$ values at vision angles (a,d) $\psi=\pi$, (b,e) $\psi=\pi/2$, and (c,f) $\psi=\pi/4$.}
    \label{fig:dynamics_supp}
\end{figure}

{\em Dynamics, Mean-Squared Displacement, and Fractional Brownian Motion --}
To better understand the dynamics of the agents, we compute their mean-squared 
displacement (MSD)
\begin{equation}
    \langle \boldsymbol{r}^2(t) \rangle = \langle |\boldsymbol{r}(t+\tau) - \boldsymbol{r}(\tau)|^2 \rangle_\tau, 
\end{equation}
where $\boldsymbol{r}(t)$ is the position vector of the particle at time $t$. The MSD curves [see extended data 
Fig.~\ref{fig:dynamics_supp}] indicate that increasing $\Delta$ or $\psi$ leads to larger scattering causing a shift of the motion from ballistic to super-diffusive. We also calculate the orientational auto-correlation function
\begin{equation}
    C(t) = \langle \boldsymbol{e}_i(t+\tau)\cdot \boldsymbol{e}_i(\tau)\rangle_\tau
\end{equation}
for different $\Delta$ at $\psi=\pi/2$ [see extended data Fig.~\ref{fig:dynamics_supp}]. 
For small $\Delta \lesssim 1$, the motion is strongly correlated, i.e. the particles 
hardly change their direction of motion as they have a strong tendency to orient and 
move toward the goal and do not scatter. For $\Delta \gtrsim 1$, the auto-correlation function $C(t)$ displays a slow
power-law decay, consistent with the super-diffusive behavior observed in the MSD $\langle \boldsymbol{r}^2(t) \rangle = K_\alpha t^{\alpha}$.

\begin{figure}[h]
    \centering
    \includegraphics[scale=0.7]{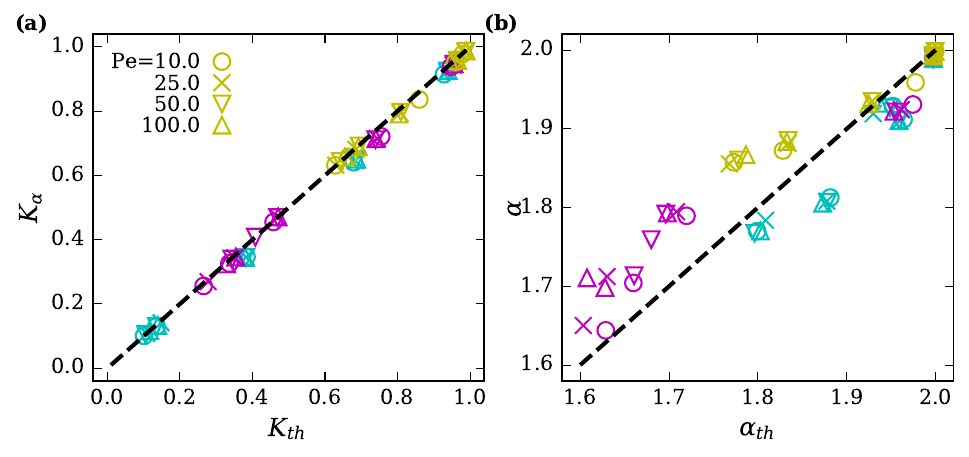}
    \caption{(a) Measured diffusion coefficient $K_\alpha$ from the MSD vs. the estimated $K_{th}=2dD_H/(\Gamma(2H+1)$ [$H$, $D_H$ estimated from $C(t)$]. 
    (b) Measured exponent $\alpha$ from the MSD vs the estimated $\alpha_{th}=2H$. 
    The different symbols correspond to different Pe number and the different 
    colours correspond to different vision angles [$\pi$ (blue), $\pi/2$ (magenta), $\pi/3$ (green)]}
    \label{fig:dynamics_supp_theory}
\end{figure}

The observed functional forms of $C(t)$ and the MSD suggest that the motion of a single agent can be described by fractional Brownian motion \cite{mandelbrot1968fractional}, where 
the velocity auto-correlation has the form [see SI]
\begin{equation}
    C(t) = \frac{dD_H}{\Gamma(2H-1)} t^{2H-2}
    \label{eg:corr}
\end{equation}
with the corresponding MSD
\begin{equation}
   \langle \boldsymbol{r}^2(t) \rangle= \frac{2dD_H}{\Gamma(2H+1)} t^{2H},
   \label{eq:msd}
\end{equation}
where $d$ is the spatial dimensionality, $H$ is the Hurst exponent and $D_H$ measures 
the strength of the coloured noise. Assuming the power law decay of $C(t)$ has the form given by Eq.~\eqref{eg:corr}, it can be checked that the coefficient $K_\alpha$ and exponent $\alpha$ measured from the MSD match the expected values $\alpha_{th}=2H$ and $K_{th} = 2dD_H/\Gamma(2H+1)$, with $H$ and $D_H$ fitted from $C(t)$[see extended data Fig.~\ref{fig:dynamics_supp_theory}]. This implies that a fractional Brownian motion of the agents well describes the scattering process. Figure~\ref{fig:dynamics}(a) shows a marginal variation in the Hurst exponent with $0.8<H<1$, consistent with the super-diffusive/ballistic motion of the agents ($\alpha = 2H$). In particular, a value $H>1/2$ indicates the long memory effect of the noise, a consequence of the goal oriented motion of the pedestrians. Notably, there is a strong decrease in $D_H$ for increasing $\Delta$ and (or) $\psi$ due to more scattering events, thus implying smaller diffusion coefficient $K_\alpha \propto D_H$, see Fig.~\ref{fig:dynamics}(b). Note that in our case the 'noise' strength of the fMB is related to the average step length taken by the agent, which decreases for increased scattering. Thus an {\em increase} in 'scattering induced noise' causes an {\em decrease} in the fBM noise, an important correspondence to keep in mind. From the analysis it can be concluded that the interactions lead to an overall decrease in the effective velocity captured via $D_H$, while the motion is still overall 'goal-oriented' i.e super-diffusive. 

Notably at $\Delta=8$, the jammed/percolated phase for $\psi=\pi$ has a larger Hurst exponent $H$ compared to the 'scattering' state at $\Delta=4$ and $\psi=\pi$, despite of nearly equal diffusion coefficient $D_H$. Here, by moving within the percolated cluster, the agents are able to achieve directed movement with larger long-time memory effects (i.e. $H$) as the agents follow the cluster that has already made its way to the exit, 
highlighting the unique dynamics of the jammed state [see Movie M3]. Qualitatively, this 
phenomenon resembles the behaviour of pedestrians joining forces to penetrate 
through highly congested areas, reflecting a collective strategy to navigate 
through crowded spaces more effectively. These results are striking, as they suggest that the complex motion involving a combination of noise, goal fixation, and vision-based steering avoidance between several other agents can be described in a mean-field manner by a coloured-noise approximation.

{\em Dependence on Inflow Rate --}
The (dimensionless) inflow rate $\Gamma$, i.e. the number of agents entering at 
the inflow region per unit time $\tau_r$, is an important parameter that determines 
the emergent behaviour in the interaction regime. Already in the much simpler 
scenario of single-file motion with volume exclusion, changing the inflow (and 
similarly the outflow) rate can lead to so-called boundary-induced phase transitions 
\cite{Krug91}. We focus on the regime of large $\Delta$ representing the case of 
strong avoidance, and study the effect of changing inflow rate on the agent motion 
for different vision angles $\psi$.

The fundamental diagram for the pedestrian movement relates flux $J=\rho\Bar{v}$ and average velocity $\bar{v}$ to the local density $\rho$, see Fig.~\ref{fig:rate}(a,b). The local density $\rho=N d_{neigh}^2/R_\text{int}^2$ in the interaction region (of radius $R_\text{int}/2$) is measured by approximating the area occupied by each agent by the average minimum separation $d_{neigh}(\Gamma,\psi)$ within the vision range $R_v$ 
[see SI]. Since the motion is largely ballistic, we approximate the average velocity as $\bar{v}= \sqrt{K_\alpha}$. Notably, we observe a collapse of the data for different vision angles onto a single master curve with the characteristic form of the  fundamental diagram, i.e.~a free-flow regime at low $\rho$ and the jammed state at high $\rho$. Even without explicit velocity adaptation in our model, we successfully replicate essential features of the fundamental diagram for pedestrian flow \cite{vanumu2017fundamental}. This implies that the model demonstrates robust properties when examined from a statistical viewpoint. Additionally, we conclude from our simulation results that the fundamental diagram holds even for different vision angles. 

\begin{figure}
    \centering
    \includegraphics[scale=0.7]{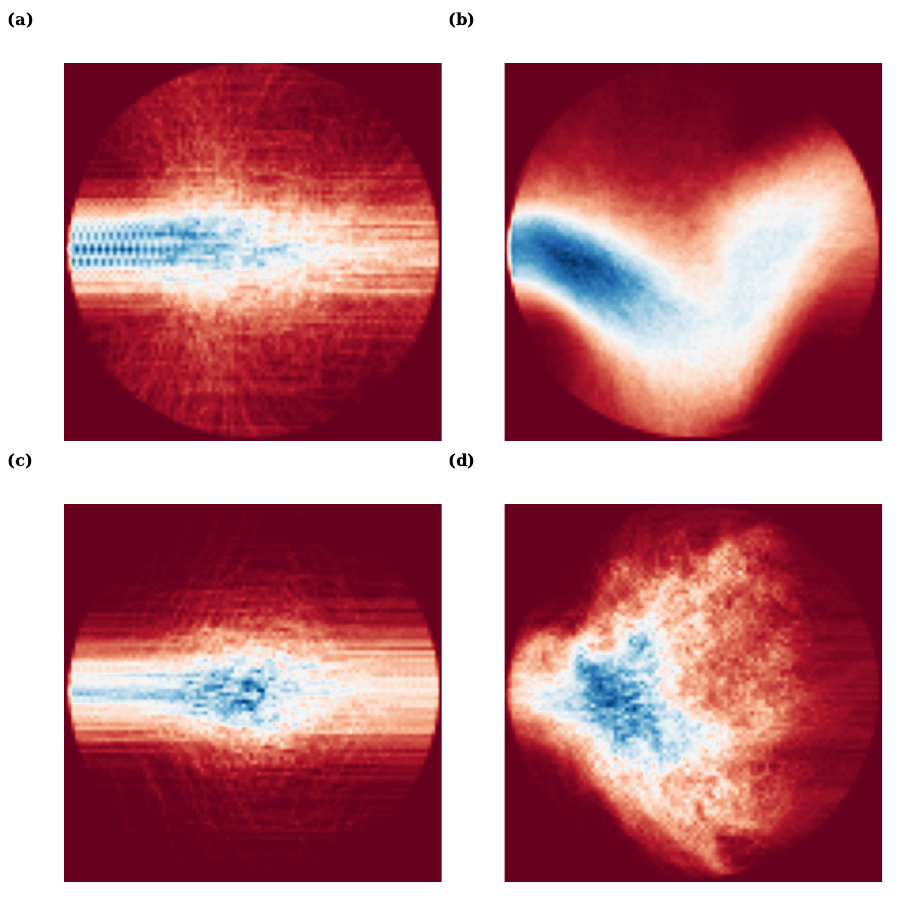}
    \caption{Local density with varying inflow for $\psi=\pi/2$, and inflow (a) $\Gamma=0.5$ and (b) $\Gamma=4$. As $\Gamma$ increases, a  rotation phase with an asymmetric density develops, as seen in (b). Local density with varying inflow for $\psi=\pi$, and inflow (a) $\Gamma=$0.4 and (c) $\Gamma=$0.6. At $\Gamma=0.4$, there is no jamming, as can be seen by the inflow and outflow density lines. However, as $\Gamma\geq 0.5$, the system enters a jammed state, characterized by the depleted outflow lines and crowding at the inflow. Here, agents enter at the left and exit at the right and the heatmap is shown for a single pedestrian stream.  }
    \label{fig:density_plots_supp}
\end{figure}

In the free-flow regime, we have a steady state and the inflow equals the outflow. 
However, at $\psi=\pi$, a jamming transition occurs around 
$\Gamma_{crit} \approx 0.5$, leading to a sudden rise in the local density 
($\rho>1.0$) [see extended data Fig.~\ref{fig:density_plots_supp}]. In this case, a large number of agents exit close to the entry due 
to overcrowding in the interaction regime and the outflow saturates. The jamming 
is triggered by the limited transport capacity in the interaction regime, and thus 
generally $\Gamma_{crit}=\Gamma_{crit}(R)$, where $R$ is the system size. This is 
notably different from boundary-induced phase transitions in one-dimensional systems 
\cite{Krug91}, which show no system-size dependence. Moreover, the monotonic decay 
of $\Bar{v}(\rho)$ suggests that the free-flow to jamming transition may be 
understood as motility-induced jamming \cite{cates2015motility}, where the repulsive
conservative potential of standard ABPs is replaced here by a vision-assisted 
steering avoidance. At smaller vision angles, the system maintains the free-flow 
regime even at large inflow, as agents allow for closer proximity [see SI], 
preventing congestion. Increasing $\rho$ (i.e. $\Gamma$) causes a decrease in the 
average velocity due to increased scattering in the interaction regime, as seen 
in Fig.~\ref{fig:rate}(b). In particular for the jammed state ($\rho\gtrsim1.0$), 
a sharp reduction in the velocity is seen. As before, the jammed state has a heightened value of $\alpha$ due to the long time persistent motion of particles within percolated clusters.

Interestingly for $\psi=\pi/2$, different movement strategies emerge as the inflow 
rate $\Gamma$ is increased. For instance, at $\Gamma = 2.0$ and $4.0$, a rotation 
state develops wherein agents follow other agents with the same goal and form a 
vortex around the center of the interaction, see Fig.~\ref{fig:rate}(c). This rotation state also allows for lower repulsion as each particle is largely aligned with the neighbors, see 
Eq.~\eqref{eq:tij2} in Methods. This motion also creates an `eye` in the centre, marked by agent 
depletion [see extended data Fig.~\ref{fig:density_plots_supp} and Movie M4], suggestive of 
traffic at a roundabout. This is consistent with the observation of several studies 
that show the stabilization of pedestrian flows at intersections in presence of an 
obstacle \cite{ondvrej2010synthetic,helbing2001self}. Thus, the self-organized 
development of the 'eye' in the centre in our simulations leads to a stabilization 
of flow. This emergent global state offers insights into discerning effective 
transport strategies contingent upon the pedestrian volume.

{\em Summary and Conclusions --} Drawing inspiration from active matter 
models in biophysics, we have introduced a new approach to simulating pedestrian 
motion, which employs a vision-based steering mechanism of agents in combination 
with goal fixation. Unlike the Social Force Model and its derivatives which employ conservative forces via potentials for obstacle avoidance and goal following, in this work we employ a local 
environment-based self-steering through torques that alter the propulsion 
direction of the agents. The overdamped limit of the Langevin equation mitigates artefacts arising from inertia effects and Newton's action-reaction principle, which is also inherent in
force-based mechanistic pedestrian models. This allows for more realistic pedestrian motion, who 'steer' their movement direction, rather than face repulsive/attractive forces, with the former navigation strategy likely dominating in low-density scenarios.  Moreover, we successfully isolate the effects of different parameters such as relative maneuverability 
$\Delta$, vision angle $\psi$, and inflow rate $\Gamma$ on the collective dynamics 
of the agents.

In the state diagram, four classes of motion patterns of semi-dense crowds are 
obtained, in which agents are weakly interacting, flocking, strongly scattering, 
and jamming. Notably, the jammed state for $\psi=\pi$ is characterized by percolating
clusters, which qualitatively resembles the behaviour of pedestrians joining forces 
to penetrate through highly congested areas. 
Despite of large differences in the global collective behaviour, the complex interplay of inter-agent interactions and goal fixation, the observed super-diffusive motion of the agents can be explained very well using a fractional Brownian motion model, featuring highly correlated noise with long memory. For increasing inflow, agents display distinct collective behaviors based on their vision angle, 
such as the development of roundabout motion at $\psi=\pi/2$ and a jammed state at $\psi=\pi$. Remarkably, the fundamental flow diagram is found to be universal for different vision angles.  

Our study lays the groundwork for more detailed modelling of pedestrian navigation 
scenarios. By introducing additional torques, e.g. related to the presence of 
boundaries, more complex scenarios such as navigation through channels, bottlenecks,
and obstacle avoidance can be studied. An important next step will be the addition 
of adaptable agent velocities, particularly relevant in high-density crowd simulations. 
Lastly, we predict the validity of the fundamental diagram for different vision angles, which would be interesting to study experimentally.


\begin{thebibliography}{39}
	\ifx \bisbn   \undefined \def \bisbn  #1{ISBN #1}\fi
	\ifx \binits  \undefined \def \binits#1{#1}\fi
	\ifx \bauthor  \undefined \def \bauthor#1{#1}\fi
	\ifx \batitle  \undefined \def \batitle#1{#1}\fi
	\ifx \bjtitle  \undefined \def \bjtitle#1{#1}\fi
	\ifx \bvolume  \undefined \def \bvolume#1{\textbf{#1}}\fi
	\ifx \byear  \undefined \def \byear#1{#1}\fi
	\ifx \bissue  \undefined \def \bissue#1{#1}\fi
	\ifx \bfpage  \undefined \def \bfpage#1{#1}\fi
	\ifx \blpage  \undefined \def \blpage #1{#1}\fi
	\ifx \burl  \undefined \def \burl#1{\textsf{#1}}\fi
	\ifx \doiurl  \undefined \def \doiurl#1{\url{https://doi.org/#1}}\fi
	\ifx \betal  \undefined \def \betal{\textit{et al.}}\fi
	\ifx \binstitute  \undefined \def \binstitute#1{#1}\fi
	\ifx \binstitutionaled  \undefined \def \binstitutionaled#1{#1}\fi
	\ifx \bctitle  \undefined \def \bctitle#1{#1}\fi
	\ifx \beditor  \undefined \def \beditor#1{#1}\fi
	\ifx \bpublisher  \undefined \def \bpublisher#1{#1}\fi
	\ifx \bbtitle  \undefined \def \bbtitle#1{#1}\fi
	\ifx \bedition  \undefined \def \bedition#1{#1}\fi
	\ifx \bseriesno  \undefined \def \bseriesno#1{#1}\fi
	\ifx \blocation  \undefined \def \blocation#1{#1}\fi
	\ifx \bsertitle  \undefined \def \bsertitle#1{#1}\fi
	\ifx \bsnm \undefined \def \bsnm#1{#1}\fi
	\ifx \bsuffix \undefined \def \bsuffix#1{#1}\fi
	\ifx \bparticle \undefined \def \bparticle#1{#1}\fi
	\ifx \barticle \undefined \def \barticle#1{#1}\fi
	\bibcommenthead
	\ifx \bconfdate \undefined \def \bconfdate #1{#1}\fi
	\ifx \botherref \undefined \def \botherref #1{#1}\fi
	\ifx \url \undefined \def \url#1{\textsf{#1}}\fi
	\ifx \bchapter \undefined \def \bchapter#1{#1}\fi
	\ifx \bbook \undefined \def \bbook#1{#1}\fi
	\ifx \bcomment \undefined \def \bcomment#1{#1}\fi
	\ifx \oauthor \undefined \def \oauthor#1{#1}\fi
	\ifx \citeauthoryear \undefined \def \citeauthoryear#1{#1}\fi
	\ifx \endbibitem  \undefined \def \endbibitem {}\fi
	\ifx \bconflocation  \undefined \def \bconflocation#1{#1}\fi
	\ifx \arxivurl  \undefined \def \arxivurl#1{\textsf{#1}}\fi
	\csname PreBibitemsHook\endcsname
	
	\bibitem[\protect\citeauthoryear{Feliciani et~al.}{2021}]{FelicianiBook}
	\begin{bbook}
		\bauthor{\bsnm{Feliciani}, \binits{C.}},
		\bauthor{\bsnm{Shimura}, \binits{K.}},
		\bauthor{\bsnm{Nishinari}, \binits{K.}}:
		\bbtitle{Introduction to Crowd Management: Managing Crowds in the Digital Era:
			Theory and Practice}.
		\bpublisher{Springer},
		\blocation{Cham, Switzerland}
		(\byear{2021})
	\end{bbook}
	\endbibitem
	
	\bibitem[\protect\citeauthoryear{Schadschneider
		et~al.}{2018}]{SchadschneiderCSTZ18}
	\begin{barticle}
		\bauthor{\bsnm{Schadschneider}, \binits{A.}},
		\bauthor{\bsnm{Chraibi}, \binits{M.}},
		\bauthor{\bsnm{Seyfried}, \binits{A.}},
		\bauthor{\bsnm{Tordeux}, \binits{A.}},
		\bauthor{\bsnm{Zhang}, \binits{J.}}:
		\batitle{Pedestrian dynamics: From empirical results to modeling}.
		\bjtitle{Crowd Dynamics, Volume 1: Theory, Models, and Safety Problems}
		\bvolume{1},
		\bfpage{63}
		(\byear{2018})
	\end{barticle}
	\endbibitem
	
	\bibitem[\protect\citeauthoryear{Corbetta and
		Toschi}{2023}]{corbetta2023physics}
	\begin{barticle}
		\bauthor{\bsnm{Corbetta}, \binits{A.}},
		\bauthor{\bsnm{Toschi}, \binits{F.}}:
		\batitle{Physics of human crowds}.
		\bjtitle{Annu. Rev. Condens. Matter Phys.}
		\bvolume{14},
		\bfpage{311}--\blpage{333}
		(\byear{2023})
	\end{barticle}
	\endbibitem
	
	\bibitem[\protect\citeauthoryear{Zhang et~al.}{2012}]{ZhangKSS12}
	\begin{botherref}
		\oauthor{\bsnm{Zhang}, \binits{J.}},
		\oauthor{\bsnm{W.W.F.Klingsch}},
		\oauthor{\bsnm{Schadschneider}, \binits{A.}},
		\oauthor{\bsnm{Seyfried}, \binits{A.}}:
		Ordering in bidirectional pedestrian flows and its influence on the fundamental
		diagram.
		JSTAT,
		02002
		(2012)
	\end{botherref}
	\endbibitem
	
	\bibitem[\protect\citeauthoryear{Feliciani and Nishinari}{2016}]{FelicianiN16b}
	\begin{barticle}
		\bauthor{\bsnm{Feliciani}, \binits{C.}},
		\bauthor{\bsnm{Nishinari}, \binits{K.}}:
		\batitle{Empirical analysis of the lane formation process in bidirectional
			pedestrian flow}.
		\bjtitle{Phys. Rev. E}
		\bvolume{94},
		\bfpage{032304}
		(\byear{2016})
	\end{barticle}
	\endbibitem
	
	\bibitem[\protect\citeauthoryear{Dong et~al.}{2017}]{dong2017analysis}
	\begin{barticle}
		\bauthor{\bsnm{Dong}, \binits{H.-R.}},
		\bauthor{\bsnm{Meng}, \binits{Q.}},
		\bauthor{\bsnm{Yao}, \binits{X.-M.}},
		\bauthor{\bsnm{Yang}, \binits{X.-X.}},
		\bauthor{\bsnm{Wang}, \binits{Q.-L.}}:
		\batitle{Analysis of dynamic features in intersecting pedestrian flow}.
		\bjtitle{Chin. Phys. B}
		\bvolume{26}(\bissue{9}),
		\bfpage{098902}
		(\byear{2017})
	\end{barticle}
	\endbibitem
	
	\bibitem[\protect\citeauthoryear{Mullick et~al.}{2022}]{mullick2022analysis}
	\begin{barticle}
		\bauthor{\bsnm{Mullick}, \binits{P.}},
		\bauthor{\bsnm{Fontaine}, \binits{S.}},
		\bauthor{\bsnm{Appert-Rolland}, \binits{C.}},
		\bauthor{\bsnm{Olivier}, \binits{A.-H.}},
		\bauthor{\bsnm{Warren}, \binits{W.H.}},
		\bauthor{\bsnm{Pettr{\'e}}, \binits{J.}}:
		\batitle{Analysis of emergent patterns in crossing flows of pedestrians reveals
			an invariant of ‘stripe’formation in human data}.
		\bjtitle{PLoS Comput. Biol.}
		\bvolume{18}(\bissue{6}),
		\bfpage{1010210}
		(\byear{2022})
	\end{barticle}
	\endbibitem
	
	\bibitem[\protect\citeauthoryear{Bode et~al.}{2019}]{BodeCH19}
	\begin{barticle}
		\bauthor{\bsnm{Bode}, \binits{N.W.F.}},
		\bauthor{\bsnm{Chraibi}, \binits{M.}},
		\bauthor{\bsnm{Holl}, \binits{S.}}:
		\batitle{The emergence of macroscopic interactions between intersecting
			pedestrian streams}.
		\bjtitle{Transp. Res. B Methodol.}
		\bvolume{119},
		\bfpage{197}
		(\byear{2019})
	\end{barticle}
	\endbibitem
	
	\bibitem[\protect\citeauthoryear{Cao et~al.}{2017}]{cao2017fundamental}
	\begin{barticle}
		\bauthor{\bsnm{Cao}, \binits{S.}},
		\bauthor{\bsnm{Seyfried}, \binits{A.}},
		\bauthor{\bsnm{Zhang}, \binits{J.}},
		\bauthor{\bsnm{Holl}, \binits{S.}},
		\bauthor{\bsnm{Song}, \binits{W.}}:
		\batitle{Fundamental diagrams for multidirectional pedestrian flows}.
		\bjtitle{Journal of Statistical Mechanics: Theory and Experiment}
		\bvolume{2017}(\bissue{3}),
		\bfpage{033404}
		(\byear{2017})
	\end{barticle}
	\endbibitem
	
	\bibitem[\protect\citeauthoryear{Lian et~al.}{2015}]{lian2015experimental}
	\begin{barticle}
		\bauthor{\bsnm{Lian}, \binits{L.}},
		\bauthor{\bsnm{Mai}, \binits{X.}},
		\bauthor{\bsnm{Song}, \binits{W.}},
		\bauthor{\bsnm{Richard}, \binits{Y.K.K.}},
		\bauthor{\bsnm{Wei}, \binits{X.}},
		\bauthor{\bsnm{Ma}, \binits{J.}}:
		\batitle{An experimental study on four-directional intersecting pedestrian
			flows}.
		\bjtitle{JSTAT}
		\bvolume{2015}(\bissue{8}),
		\bfpage{08024}
		(\byear{2015})
	\end{barticle}
	\endbibitem
	
	\bibitem[\protect\citeauthoryear{Sun et~al.}{2018}]{SunHGQC18}
	\begin{barticle}
		\bauthor{\bsnm{Sun}, \binits{L.}},
		\bauthor{\bsnm{Hao}, \binits{S.}},
		\bauthor{\bsnm{Gong}, \binits{Q.}},
		\bauthor{\bsnm{Qiu}, \binits{S.}},
		\bauthor{\bsnm{Chen}, \binits{Y.}}:
		\batitle{Pedestrian roundabout improvement strategy in subway stations}.
		\bjtitle{Transport}
		\bvolume{171},
		\bfpage{1600073}
		(\byear{2018})
	\end{barticle}
	\endbibitem
	
	\bibitem[\protect\citeauthoryear{Xiao et~al.}{2019}]{xiao2019investigation}
	\begin{barticle}
		\bauthor{\bsnm{Xiao}, \binits{Y.}},
		\bauthor{\bsnm{Gao}, \binits{Z.}},
		\bauthor{\bsnm{Jiang}, \binits{R.}},
		\bauthor{\bsnm{Li}, \binits{X.}},
		\bauthor{\bsnm{Qu}, \binits{Y.}},
		\bauthor{\bsnm{Huang}, \binits{Q.}}:
		\batitle{Investigation of pedestrian dynamics in circle antipode experiments:
			Analysis and model evaluation with macroscopic indexes}.
		\bjtitle{Transp. Res. Part C Emerg.}
		\bvolume{103},
		\bfpage{174}--\blpage{193}
		(\byear{2019})
	\end{barticle}
	\endbibitem
	
	\bibitem[\protect\citeauthoryear{Hu et~al.}{2019}]{hu2019experimental}
	\begin{barticle}
		\bauthor{\bsnm{Hu}, \binits{Y.}},
		\bauthor{\bsnm{Zhang}, \binits{J.}},
		\bauthor{\bsnm{Song}, \binits{W.}}:
		\batitle{Experimental study on the movement strategies of individuals in
			multidirectional flows}.
		\bjtitle{Physica A: Statistical Mechanics and its Applications}
		\bvolume{534},
		\bfpage{122046}
		(\byear{2019})
	\end{barticle}
	\endbibitem
	
	\bibitem[\protect\citeauthoryear{Martinez-Gil et~al.}{2017}]{MartinezGil2017}
	\begin{barticle}
		\bauthor{\bsnm{Martinez-Gil}, \binits{F.}},
		\bauthor{\bsnm{Lozano}, \binits{M.}},
		\bauthor{\bsnm{Garcia-Fernandez}, \binits{I.}},
		\bauthor{\bsnm{Fernandez}, \binits{F.}}:
		\batitle{Modeling, evaluation, and scale on artificial pedestrians: a
			literature review}.
		\bjtitle{ACM Computing Surveys (CSUR)}
		\bvolume{50},
		\bfpage{72}
		(\byear{2017})
	\end{barticle}
	\endbibitem
	
	\bibitem[\protect\citeauthoryear{Maury and Faure}{2019}]{MauryBook}
	\begin{bbook}
		\bauthor{\bsnm{Maury}, \binits{B.}},
		\bauthor{\bsnm{Faure}, \binits{S.}}:
		\bbtitle{Crowds in Equations}.
		\bpublisher{World Scientific},
		\blocation{London, Singapore}
		(\byear{2019})
	\end{bbook}
	\endbibitem
	
	\bibitem[\protect\citeauthoryear{Boltes et~al.}{2013}]{BoltesZS13}
	\begin{barticle}
		\bauthor{\bsnm{Boltes}, \binits{M.}},
		\bauthor{\bsnm{Zhang}, \binits{J.}},
		\bauthor{\bsnm{Seyfried}, \binits{A.}}:
		\batitle{Analysis of crowd dynamics with laboratory experiments}.
		\bjtitle{The International Series in Video Computing}
		\bvolume{11},
		\bfpage{67}
		(\byear{2013})
	\end{barticle}
	\endbibitem
	
	\bibitem[\protect\citeauthoryear{Gompper et~al.}{2020}]{gompper20202020}
	\begin{barticle}
		\bauthor{\bsnm{Gompper}, \binits{G.}},
		\bauthor{\bsnm{Winkler}, \binits{R.G.}},
		\bauthor{\bsnm{Speck}, \binits{T.}},
		\bauthor{\bsnm{Solon}, \binits{A.}},
		\bauthor{\bsnm{Nardini}, \binits{C.}},
		\bauthor{\bsnm{Peruani}, \binits{F.}},
		\bauthor{\bsnm{L{\"o}wen}, \binits{H.}},
		\bauthor{\bsnm{Golestanian}, \binits{R.}},
		\bauthor{\bsnm{Kaupp}, \binits{U.B.}},
		\bauthor{\bsnm{Alvarez}, \binits{L.}}, \betal:
		\batitle{The 2020 motile active matter roadmap}.
		\bjtitle{J. Condens. Matter Phys.}
		\bvolume{32}(\bissue{19}),
		\bfpage{193001}
		(\byear{2020})
	\end{barticle}
	\endbibitem
	
	\bibitem[\protect\citeauthoryear{Fily et~al.}{2014}]{fily2014dynamics}
	\begin{barticle}
		\bauthor{\bsnm{Fily}, \binits{Y.}},
		\bauthor{\bsnm{Baskaran}, \binits{A.}},
		\bauthor{\bsnm{Hagan}, \binits{M.F.}}:
		\batitle{Dynamics of self-propelled particles under strong confinement}.
		\bjtitle{Soft matter}
		\bvolume{10}(\bissue{30}),
		\bfpage{5609}--\blpage{5617}
		(\byear{2014})
	\end{barticle}
	\endbibitem
	
	\bibitem[\protect\citeauthoryear{Cates and Tailleur}{2015}]{cates2015motility}
	\begin{barticle}
		\bauthor{\bsnm{Cates}, \binits{M.E.}},
		\bauthor{\bsnm{Tailleur}, \binits{J.}}:
		\batitle{Motility-induced phase separation}.
		\bjtitle{Annu. Rev. Condens. Matter Phys.}
		\bvolume{6}(\bissue{1}),
		\bfpage{219}--\blpage{244}
		(\byear{2015})
	\end{barticle}
	\endbibitem
	
	\bibitem[\protect\citeauthoryear{Iyer et~al.}{2023}]{iyer2023dynamics}
	\begin{barticle}
		\bauthor{\bsnm{Iyer}, \binits{P.}},
		\bauthor{\bsnm{Winkler}, \binits{R.G.}},
		\bauthor{\bsnm{Fedosov}, \binits{D.A.}},
		\bauthor{\bsnm{Gompper}, \binits{G.}}:
		\batitle{Dynamics and phase separation of active {B}rownian particles on curved
			surfaces and in porous media}.
		\bjtitle{Phys. Rev. Res.}
		\bvolume{5}(\bissue{3}),
		\bfpage{033054}
		(\byear{2023})
	\end{barticle}
	\endbibitem
	
	\bibitem[\protect\citeauthoryear{Caprini et~al.}{2022}]{caprini2022dynamics}
	\begin{barticle}
		\bauthor{\bsnm{Caprini}, \binits{L.}},
		\bauthor{\bsnm{Marconi}, \binits{U.M.B.}},
		\bauthor{\bsnm{Wittmann}, \binits{R.}},
		\bauthor{\bsnm{L{\"o}wen}, \binits{H.}}:
		\batitle{Dynamics of active particles with space-dependent swim velocity}.
		\bjtitle{Soft Matter}
		\bvolume{18}(\bissue{7}),
		\bfpage{1412}--\blpage{1422}
		(\byear{2022})
	\end{barticle}
	\endbibitem
	
	\bibitem[\protect\citeauthoryear{Couzin et~al.}{2005}]{Couzin_2005_nature}
	\begin{barticle}
		\bauthor{\bsnm{Couzin}, \binits{I.D.}},
		\bauthor{\bsnm{Krause}, \binits{J.}},
		\bauthor{\bsnm{Franks}, \binits{N.R.}},
		\bauthor{\bsnm{Levin}, \binits{S.A.}}:
		\batitle{Effective leadership and decision-making in animal groups on the
			move}.
		\bjtitle{Nature}
		\bvolume{433}(\bissue{7025}),
		\bfpage{513}--\blpage{516}
		(\byear{2005})
	\end{barticle}
	\endbibitem
	
	\bibitem[\protect\citeauthoryear{Barberis and
		Peruani}{2016}]{barberis_2016_PRL}
	\begin{barticle}
		\bauthor{\bsnm{Barberis}, \binits{L.}},
		\bauthor{\bsnm{Peruani}, \binits{F.}}:
		\batitle{Large-scale patterns in a minimal cognitive flocking model: Incidental
			leaders, nematic patterns, and aggregates}.
		\bjtitle{Phys. Rev. Lett.}
		\bvolume{117},
		\bfpage{248001}
		(\byear{2016})
	\end{barticle}
	\endbibitem
	
	\bibitem[\protect\citeauthoryear{Negi et~al.}{2022}]{negi2022emergent}
	\begin{barticle}
		\bauthor{\bsnm{Negi}, \binits{R.S.}},
		\bauthor{\bsnm{Winkler}, \binits{R.G.}},
		\bauthor{\bsnm{Gompper}, \binits{G.}}:
		\batitle{Emergent collective behavior of active {B}rownian particles with
			visual perception}.
		\bjtitle{Soft Matter}
		\bvolume{18}(\bissue{33}),
		\bfpage{6167}--\blpage{6178}
		(\byear{2022})
	\end{barticle}
	\endbibitem
	
	\bibitem[\protect\citeauthoryear{Negi et~al.}{2024}]{negi2024collective}
	\begin{barticle}
		\bauthor{\bsnm{Negi}, \binits{R.S.}},
		\bauthor{\bsnm{Winkler}, \binits{R.G.}},
		\bauthor{\bsnm{Gompper}, \binits{G.}}:
		\batitle{Collective behavior of self-steering active particles with velocity
			alignment and visual perception}.
		\bjtitle{Phys. Rev. Res.}
		\bvolume{6}(\bissue{1}),
		\bfpage{013118}
		(\byear{2024})
	\end{barticle}
	\endbibitem
	
	\bibitem[\protect\citeauthoryear{Ond{\v{r}}ej
		et~al.}{2010}]{ondvrej2010synthetic}
	\begin{barticle}
		\bauthor{\bsnm{Ond{\v{r}}ej}, \binits{J.}},
		\bauthor{\bsnm{Pettr{\'e}}, \binits{J.}},
		\bauthor{\bsnm{Olivier}, \binits{A.-H.}},
		\bauthor{\bsnm{Donikian}, \binits{S.}}:
		\batitle{A synthetic-vision based steering approach for crowd simulation}.
		\bjtitle{ACM Transactions on Graphics (TOG)}
		\bvolume{29}(\bissue{4}),
		\bfpage{1}--\blpage{9}
		(\byear{2010})
	\end{barticle}
	\endbibitem
	
	\bibitem[\protect\citeauthoryear{Wirth et~al.}{2023}]{wirth2023neighborhood}
	\begin{barticle}
		\bauthor{\bsnm{Wirth}, \binits{T.D.}},
		\bauthor{\bsnm{Dachner}, \binits{G.C.}},
		\bauthor{\bsnm{Rio}, \binits{K.W.}},
		\bauthor{\bsnm{Warren}, \binits{W.H.}}:
		\batitle{Is the neighborhood of interaction in human crowds metric,
			topological, or visual?}
		\bjtitle{PNAS Nexus}
		\bvolume{2}(\bissue{5}),
		\bfpage{118}
		(\byear{2023})
	\end{barticle}
	\endbibitem
	
	\bibitem[\protect\citeauthoryear{Zhang et~al.}{2022}]{zhang2022hdrlm3d}
	\begin{barticle}
		\bauthor{\bsnm{Zhang}, \binits{D.}},
		\bauthor{\bsnm{Li}, \binits{W.}},
		\bauthor{\bsnm{Gong}, \binits{J.}},
		\bauthor{\bsnm{Huang}, \binits{L.}},
		\bauthor{\bsnm{Zhang}, \binits{G.}},
		\bauthor{\bsnm{Shen}, \binits{S.}},
		\bauthor{\bsnm{Liu}, \binits{J.}},
		\bauthor{\bsnm{Ma}, \binits{H.}}:
		\batitle{Hdrlm3d: A deep reinforcement learning-based model with human-like
			perceptron and policy for crowd evacuation in 3d environments}.
		\bjtitle{ISPRS International Journal of Geo-Information}
		\bvolume{11}(\bissue{4}),
		\bfpage{255}
		(\byear{2022})
	\end{barticle}
	\endbibitem
	
	\bibitem[\protect\citeauthoryear{Dachner et~al.}{2022}]{dachner2022visual}
	\begin{barticle}
		\bauthor{\bsnm{Dachner}, \binits{G.C.}},
		\bauthor{\bsnm{Wirth}, \binits{T.D.}},
		\bauthor{\bsnm{Richmond}, \binits{E.}},
		\bauthor{\bsnm{Warren}, \binits{W.H.}}:
		\batitle{The visual coupling between neighbours explains local interactions
			underlying human ‘flocking'}.
		\bjtitle{Proc. R. Soc. B}
		\bvolume{289}(\bissue{1970}),
		\bfpage{20212089}
		(\byear{2022})
	\end{barticle}
	\endbibitem
	
	\bibitem[\protect\citeauthoryear{Negi et~al.}{2024}]{negi2024controlling}
	\begin{barticle}
		\bauthor{\bsnm{Negi}, \binits{R.S.}},
		\bauthor{\bsnm{Iyer}, \binits{P.}},
		\bauthor{\bsnm{Gompper}, \binits{G.}}:
		\batitle{Controlling inter-particle distances in crowds of motile, cognitive,
			active particles}.
		\bjtitle{Sci. Rep.}
		\bvolume{14}(\bissue{1}),
		\bfpage{9443}
		(\byear{2024})
	\end{barticle}
	\endbibitem
	
	\bibitem[\protect\citeauthoryear{Karamouzas
		et~al.}{2014}]{karamouzas2014universal}
	\begin{barticle}
		\bauthor{\bsnm{Karamouzas}, \binits{I.}},
		\bauthor{\bsnm{Skinner}, \binits{B.}},
		\bauthor{\bsnm{Guy}, \binits{S.J.}}:
		\batitle{Universal power law governing pedestrian interactions}.
		\bjtitle{Phys. Rev. Lett.}
		\bvolume{113}(\bissue{23}),
		\bfpage{238701}
		(\byear{2014})
	\end{barticle}
	\endbibitem
	
	\bibitem[\protect\citeauthoryear{Cordes et~al.}{2023}]{CordesCTS23}
	\begin{barticle}
		\bauthor{\bsnm{Cordes}, \binits{J.}},
		\bauthor{\bsnm{Chraibi}, \binits{M.}},
		\bauthor{\bsnm{Tordeux}, \binits{A.}},
		\bauthor{\bsnm{Schadschneider}, \binits{A.}}:
		\batitle{Single-file pedestrian dynamics: a review of agent-following models}.
		\bjtitle{Crowd Dynamics}
		\bvolume{4},
		\bfpage{143}
		(\byear{2023})
	\end{barticle}
	\endbibitem
	
	\bibitem[\protect\citeauthoryear{}{}]{supp}
	\begin{botherref}
		Supplementary Material.
		\url{URL_will_be_inserted_by_publisher}
	\end{botherref}
	\endbibitem
	
	\bibitem[\protect\citeauthoryear{Helbing et~al.}{2001}]{helbing2001self}
	\begin{barticle}
		\bauthor{\bsnm{Helbing}, \binits{D.}},
		\bauthor{\bsnm{Moln{\'a}r}, \binits{P.}},
		\bauthor{\bsnm{Farkas}, \binits{I.J.}},
		\bauthor{\bsnm{Bolay}, \binits{K.}}:
		\batitle{Self-organizing pedestrian movement}.
		\bjtitle{Environment and Planning B: Planning and Design}
		\bvolume{28}(\bissue{3}),
		\bfpage{361}--\blpage{383}
		(\byear{2001})
	\end{barticle}
	\endbibitem
	
	\bibitem[\protect\citeauthoryear{Helbing et~al.}{2005}]{helbing2005self}
	\begin{barticle}
		\bauthor{\bsnm{Helbing}, \binits{D.}},
		\bauthor{\bsnm{Buzna}, \binits{L.}},
		\bauthor{\bsnm{Johansson}, \binits{A.}},
		\bauthor{\bsnm{Werner}, \binits{T.}}:
		\batitle{Self-organized pedestrian crowd dynamics: Experiments, simulations,
			and design solutions}.
		\bjtitle{Transp. Sci.}
		\bvolume{39}(\bissue{1}),
		\bfpage{1}--\blpage{24}
		(\byear{2005})
	\end{barticle}
	\endbibitem
	
	\bibitem[\protect\citeauthoryear{Stauffer and
		Aharony}{1994}]{stauffer2018introduction}
	\begin{bbook}
		\bauthor{\bsnm{Stauffer}, \binits{D.}},
		\bauthor{\bsnm{Aharony}, \binits{A.}}:
		\bbtitle{Introduction to Percolation Theory (revised 2nd Edition)}.
		\bpublisher{CRC Press},
		\blocation{London}
		(\byear{1994})
	\end{bbook}
	\endbibitem
	
	\bibitem[\protect\citeauthoryear{Mandelbrot and
		Van~Ness}{1968}]{mandelbrot1968fractional}
	\begin{barticle}
		\bauthor{\bsnm{Mandelbrot}, \binits{B.B.}},
		\bauthor{\bsnm{Van~Ness}, \binits{J.W.}}:
		\batitle{Fractional {B}rownian motions, fractional noises and applications}.
		\bjtitle{SIAM Review}
		\bvolume{10}(\bissue{4}),
		\bfpage{422}--\blpage{437}
		(\byear{1968})
	\end{barticle}
	\endbibitem
	
	\bibitem[\protect\citeauthoryear{Krug}{1991}]{Krug91}
	\begin{barticle}
		\bauthor{\bsnm{Krug}, \binits{J.}}:
		\batitle{Boundary-induced phase transitions in driven diffusive systems}.
		\bjtitle{Phys. Rev. Lett.}
		\bvolume{67},
		\bfpage{1882}
		(\byear{1991})
	\end{barticle}
	\endbibitem
	
	\bibitem[\protect\citeauthoryear{Vanumu et~al.}{2017}]{vanumu2017fundamental}
	\begin{barticle}
		\bauthor{\bsnm{Vanumu}, \binits{L.D.}},
		\bauthor{\bsnm{Ramachandra~Rao}, \binits{K.}},
		\bauthor{\bsnm{Tiwari}, \binits{G.}}:
		\batitle{Fundamental diagrams of pedestrian flow characteristics: A review}.
		\bjtitle{Eur. Transp. Res. Rev.}
		\bvolume{9},
		\bfpage{1}--\blpage{13}
		(\byear{2017})
	\end{barticle}
	\endbibitem
	
\end{thebibliography}

\newpage


\section{Methods}\label{sec11}

\emph{Model details}
We model pedestrians moving in a three-way crossing, as shown in Fig.~\ref{fig:schematic}(a), with an interaction zone of radius $R_\text{int}$. Each agent $i$ is associated with a type $t_i$, which encodes their goal to reach the opposite side of the crossing. There are three pedestrian streams and correspondingly three agent types. The pedestrians are here modelled as two-dimensional intelligent active Brownian particles (iABPs) which experience a propulsion force $f_p$ acting along their orientation vector $\mathbf{e}_i$. Any individual variability is incorporated as noise in the equation of motion which specifies the dynamics of the position $\mathbf{r}_i$ of an iAPB:
\begin{equation}
	m \ddot{\mathbf{r}}_i =f_p\mathbf{e}_i-\gamma \dot{\mathbf{r}}_i, 
\end{equation}
where $m$ is the agent mass and $\gamma$ is the friction coefficient. The dynamics of the orientation vector $\mathbf{e}_i$ is determined by 
\begin{equation}
  \dot{\boldsymbol{e}}_i =\sqrt{2 (d-1) D_r}\boldsymbol{\Lambda}_i + \Omega \mathbf{M}_\text{vis} + K \mathbf{M}_\text{goal}, 
\end{equation}
(see also Eq.~\eqref{eq:steering}). 
The noise acts perpendicular to the direction of motion, so that
\begin{equation}
   \Lambda_i =\boldsymbol{\zeta}_i \times \boldsymbol{e}_i, 
\end{equation}
where $\boldsymbol{\zeta}_i$ is a Gaussian and Markovian random process with 
$\langle\boldsymbol{\zeta}_i(t)\rangle=0$ and $ \langle\boldsymbol{\zeta}_i(t)\cdot \boldsymbol{\zeta}_j(t')\rangle = \delta_{ij}\delta(t-t')$. The agents avoid collisions with each other via 'vision-assisted' reorientation of their propulsion direction, which is described by the torque \cite{negi2022emergent}
\begin{equation}
   \mathbf{M}_\text{vis} = \frac{-1}{N_i}\sum_{j\in VC} T_{ij}\left[\boldsymbol{e}_i \times \left (\frac{\boldsymbol{r_{ij}}}{|\boldsymbol{r_{ij}}|}\times \boldsymbol{e}_i \right)\right]  , 
   \label{eq:vis2}
\end{equation}
where $\boldsymbol{r_{ij}}= \boldsymbol{r_{j}}-\boldsymbol{r_{i}}$ is the displacement vector between particle $i$ and particle $j$, and $T_{ij}$ is a weight factor,
\begin{equation}
	 T_{ij}=e^{(-|\boldsymbol{r_{ij}}|/R_0)} [3-\boldsymbol{e}_i\cdot \boldsymbol{e}_j]/4.
	 	\label{eq:tij2}
\end{equation}
which increases the relative importance of avoiding agents moving 'head-on' towards each other ($\boldsymbol{e}_i\cdot \boldsymbol{e}_j=-1$) as opposed to co-moving agents (i.e. $\boldsymbol{e}_i\cdot \boldsymbol{e}_j=1$) by a factor $1/2$ \cite{karamouzas2014universal}. Lastly, $N_i=\sum_{j\in VC} T_{ij}$ is the normalization factor. The exponential distance dependence in Eq.~\eqref{eq:tij2} limits the range of the interaction, such that for high density of agents the effective vision range is $R_0$. The sum is over all particles $j$ that are in the vision range $VC$ of the agent $i$, with
\begin{equation}
    VC = \left\{\,j \mid \frac{\boldsymbol{r_{ij}}}{|\boldsymbol{r_{ij}}|}\cdot e_i \geq \cos\psi \text{ and } |\boldsymbol{r_{ij}}|<R_v \,\right\}
\end{equation}
where $\psi$ is the vision angle and $R_v>R_0$ the 'full' vision range. Steering toward the goal is determined by the torque
\begin{equation}
   M_\text{goal} =\left[ \boldsymbol{e}_i \times \left (\boldsymbol{\hat{d}}(t_i)\times \boldsymbol{e}_i \right)\right]  , 
   \label{eq:lang_orient2}
\end{equation}
where the unit vector $\boldsymbol{\hat{d}}(t_i)$ is direction toward the goal, 
with which particle $i$ attempts to align [see Fig.~\ref{fig:schematic}(a)]. We define the relative maneuverability $\Delta=\Omega/K$, 
which measures the relative strength of visual avoidance to target alignment.  The combined effect of alignment and maneuverability is shown in Fig.~\ref{fig:schematic}(c-f). Pedestrians navigate their movement paths based on visual cues to avoid collisions with others, by adjusting their propulsion direction. For larger relative  maneuverability $\Delta$, the agents make sharper turns, see Fig.~\ref{fig:schematic}(d,f). Importantly, the agents' vision-based interactions are non-reciprocal for vision angle $\psi<\pi$, see Fig.~\ref{fig:schematic}(c,d).  Here the trailing agent notices the leading agent, but not vice versa. 

All influxes are spaced at the same angular distance from each other and agents enter with frequency $v_0/R_0$. At each inflow, the start position $\boldsymbol{r}_0=(x_0,y_0)$ of the incoming agent on the circle is determined by first sampling a number $x'_0$ from a normal distribution $x'_0\sim \mathcal{N}(0,\sigma^2)$ with zero mean and standard deviation $\sigma=\pi R_\text{int}/18$ to generate the intermediate point $\boldsymbol{r^{'}}_0=(x^{'}_0,y^{'}_0)$ using $x^{'2}_0+y^{'2}_0 = R_\text{int}^2$. The desired point is then generated by a rotation, $\boldsymbol{r}_0=\textbf{R}_{\theta}\boldsymbol{r^{'}_0}$, where $\theta=0,2\pi/3,-2\pi/3$ for the red, green and blue streams respectively. The value of $\sigma$ determines the \textit{approximate} interaction radius $R_\text{eff}$, via the relation $N_\text{stream}(6\sigma)\simeq 2\pi R_\text{eff}$. For our choice of  $\sigma=\pi R_\text{int}/18$, this results in an effective minimum interaction zone of radius $R_\text{eff}=R_\text{int}/2$. An agent crossing the boundary of the interaction zone at any point is removed from the simulation (absorbing boundary). The activity of the agents is described by the dimensionless P{\'e}clet number $\text{Pe} = f_p/(\gamma R_0 D_r)=v_0\tau_r/R_0$, where $\tau_r=1/D_r$ is the rotational diffusion time and $v_0=f_p/\gamma$ is the agent velocity. We operate in the over-damped limit, i.e. $m/\gamma\ll 1$ so that inertial effects are negligible and $f_p\mathbf{e}_i\approx \gamma \dot{\mathbf{r}}_i$.  All lengths are measured in units of $R_0$, time in units of $\tau_r$. The goal fixation is set to $K=8D_r$ and the vision range $R_v=4R_0$. The parameters $\Delta$, activity Pe, and vision angle $\psi$ are varied. The inflow rate $\Gamma$ measures the number of agents entering the interaction circle at each inflow per unit time ($\tau_r$). Excluded volume effects are neglected, as we operate in the limit of semi-dense crowds, i.e. $\sigma \ll R_0$, where $\sigma$ is the volume-exclusion radius of an individual pedestrian.

\backmatter

\bmhead{Supplementary information} 
Supplementary text and movie captions.



\clearpage
\section{Extended Data}

\clearpage
\clearpage



\end{document}